\documentclass{elsart}
\usepackage{graphicx,amsmath,amssymb,bm}
\usepackage{amsfonts,amssymb,amscd,amsmath}
\usepackage{graphicx}

\newcommand\fverb{\setbox\pippobox=\hbox\bgroup\verb}
\newcommand\fverbdo{\egroup\medskip\noindent%
                        \fbox{\unhbox\pippobox}\ }
\newcommand\fverbit{\egroup\item[\fbox{\unhbox\pippobox}]}

 \newcommand\beq{\begin{equation}}
 \newcommand\eeq{\end{equation}}
 \newcommand\beqn{\begin{eqnarray}}
 \newcommand\eeqn{\end{eqnarray}}

\def\GeV{\,\mbox{GeV}}

\def\GeV{\,\mbox{GeV}}

\begin{document}
\begin{frontmatter}
\title{Cronin effect for protons and pions in high-energy pA collisions}

\author{A. H. Rezaeian and }  
\author{Zhun Lu}

\vspace{0.1in}

\address{Departamento de F\'\i sica y Centro de Estudios Subat\'omicos, Universidad
T\'ecnica Federico Santa Mar\'\i a, Casilla 110-V, Valpara\'\i so,
Chile}

\vskip 0.1in



\begin{abstract}

\noindent
Pions and protons production cross-sections are analyzed in
proton-proton and proton-nucleus collisions at the RHIC energy at
midrapidity. We employ the pQCD factorization scheme supplemented with
the color-dipole formalism to investigate the Cronin effect. We
calculate the broadening in the color-dipole approach for different
centralities. Our main goal is to investigate, in a parameter-free
manner within a unified framework, how much of the cronin effect for
both pions and baryons comes from the transverse momentum broadening
due to initial partons multi-scatterings. We conclude that final-state
effects in pA collisions are important. Uncertainties in nuclear
shadowing of various parton distributions and parton fragmentation
functions are also discussed.
\vspace{0.5cm}

\end{abstract}
\end{frontmatter}



\section{Introduction}
It is believed that $p+A$ collisions provide a decisive testing ground
to distinguish between the initial- and final-state (plasma) effects in $A+A$ collisions and can be
used as a baseline for jet-quenching models.  The Cronin effect \cite{cron}(which
is generally associated with the ratio of $p+A$ and $p+p$ cross sections,
scaled by the number of collisions) in high-energy collisions has been
the subject of renewed interest in recent years
\cite{rev,wang,ktc2,boris1,fin1,cgc,break,lhcf}. In order to pin down the
role of parton energy loss effects in heavy ion collisions, a precise
and firm understanding of the underlying dynamics of the Cronin and
the shadowing effects in $p+A$ collisions is indispensable.

There have been two very different approaches to explain the Cronin
effect in $p+A$ collisions: the initial-state effects \cite{cron,rev}
due to the broadening of the parton transverse momentum in the
initial-state where the fragmentation of hard partons is assumed to
occur outside the cold medium, and final-state effects \cite{fin1} due
to the recombination of soft and shower partons in the final-state.
To understand the role of initial-state effects in the $p+A$ reactions,
one should also understand the broadening of transverse momentum of a
projectile parton propagating and interacting with a nuclear medium.
A first principle calculation of the broadening of transverse momentum
of partons is very complicated which involves apparently both soft and
hard interactions.  A common practice involves fitting to a given
experiment and then extrapolation to another reaction
\cite{wang}. However, such an approach is less reliable since
broadening is not universal and depends on kinematics and reaction.

Here, we employ the pQCD factorization scheme supplemented with the
color-dipole formalism to investigate the Cronin effect for identified
hadrons. The broadening is calculated in a parameter-free way in the
color-dipole approach. Our modest goal is to investigate the role of initial-state
effects in the observed Cronin ratio for both protons and pions in $p+A$ collisions and to learn how much of the effect comes from the
broadening due to initial partons multi-scatterings.
One of the uncertainties in such a calculation from the outset is due
to our lack of knowledge about the baryon fragmentation functions. We
investigate the implication of the uncertainties among various nuclear
parton distributions and parton fragmentation functions and the role of
parton primordial transverse momentum on the Cronin effect.

One of the stunning experimental observations in both $p+A$ and $A+A$
collisions has been the much larger magnitude of the Cronin effect on
baryons compared to mesons at the intermediate transverse momentum,
the so-called baryon/meson anomaly in high energy nuclear collisions
\cite{star-data,star}. There have been different approaches to
understand this phenomenon \cite{fin1,fin2}. But a clear explanation
is still lacking. Here, we compute the protons/pions ratio in $p+p$
and $p+A$ collisions. We also address if the underlying production mechanism of baryons and
mesons are the same.


 
This paper is organized as follows: In Sec 2, we discuss hadron
production in $p+p$ collisions. In Sec. 3, we introduce the main elements of our
computation hadron spectra in $p+A$ collisions. In particular, we discuss how to
compute broadening in the color-dipole formalism. Sec. 4 summarizes our work.

\section{Particle production in $p+p$ collisions}

High-$p_{T}$ inclusive particle production in hadron scattering may be
described by collinear pQCD factorization. However, in the derivation of the collinear factorization theorems, one
applies to the hard scattering the approximation that the transverse
momentum of the incoming parton can be neglected with respect to the
transverse momentum generated in the scattering, and one also neglects
the transverse momentum generated in the fragmentation. When the
transverse momentum of the incoming partons is in the order of hard
scale $Q$, the errors in the collinear approximations can be
compensated by a correct treatment of higher order corrections to the
hard scattering. However, for low transverse momentum compared with
$Q$, in collinear factorization there is no precise compensation from
higher order correction and one should not neglect the transverse
momentum of the incoming partons.

Moreover, it is well known that the curvature of the hadron spectrum
can be corrected by considering intrinsic transverse momentum $k_T$
for the colliding partons. It was observed that significant parton
intrinsic transverse momentum is needed for describing data of
Drell-Yan dilepton production~\cite{hom76}, direct photon
production~\cite{huston95,apanasevich98} and heavy quark
production~\cite{mangano}. The origin of such a large primordial
transverse momentum still needs a clear explanation. Nevertheless,
some higher order pQCD corrections and soft gluon radiation
corrections are effectively embodied in the primordial transverse
momentum distribution. There are also some spin-dependent effects
measured by experiments, such as single spin asymmetry, which cannot
be explained by collinear factorization in all orders of QCD. The only
way to produce these non-zero effects is with the presence of parton
intrinsic transverse momentum \cite{spin} or by considering the
higher-twist contributions. 

There exists many ways of incorporating the parton intrinsic
transverse momentum phenomenologically, and we choose for simplicity a
$k_T$ smearing of the cross section to approximate this effect via
unintegrated parton distributions. This pQCD-improved parton model
based on factorization should not be mixed with the collinear
factorization. Notice that we do not aim at comparing different
types of factorizations to claim which one is better or not. Each of
them can be applied to some specific processes and specific
kinematical region. Here, we prefer to use a kind of
$k_T$-factorization which is flexible enough in the case of $p+A$ collisions.

By assuming the validation of
the factorization in high-$p_T$ particle production, and considering
the intrinsic transverse momenta of the initial partons, the
differential cross section in $p+p$ collision can be written
as \cite{owe},
\begin{eqnarray}
\frac{d \sigma^{pp\to h+X}}{dy d^2 p_T} &=&  \sum_{ijnl}\int dx_i dx_j d^2 k_{iT}d^2 k_{jT}
f_{i/p}(x_i,Q^2) \mathcal{G}_p(k_{iT},Q^2) \nonumber\\
&\times &f_{j/p}(x_j,Q^2)\mathcal{G}_p(k_{jT},Q^2)
K\,\frac{d
\sigma}{d \hat{t}}(ij \to nl) \, \frac{D_{h/n}(z_n,Q^2)}{\pi z_n}, \label{pptopi}
\end{eqnarray}
 where the $K$ factor which is in general $\sqrt{s}$ and scale
 dependent accounts for the contribution of the NLO corrections (for
 the Cronin ratio which is the main subject of this paper, the $K$
 factor drops out and is not more important). $f_{i/p}(x_i,Q^2)$ is the parton
 distribution functions (PDF) of the colliding protons, which depend
 on the light-cone momentum fractions $x_i$ and the hard scale
 $Q$. The function $D_{h/n}(z_n,Q^2)$ is the fragmentation function
 (FF) of parton $n$ to the final hadron $h$ with a momentum fraction
 $z_n$. The differential cross section $\frac{d \sigma}{d \hat{t}}(ij
 \to nl)$ of the hard process $a (k_i) + b (k_j) \to c (k_n) + d
 (k_l)$ can be calculated perturbatively(in the running coupling
 $\alpha_s(Q^2)$ with scale parameter $\Lambda$ equal to pion mass),
 in terms of the following Mandelstam variables
\begin{eqnarray}
\hat{s}  =  (k_i+k_j)^2 &=& x_ix_j s+\frac{k_{iT}^2 k_{jT}^2}{x_i x_j
s}-2
\boldsymbol{k}_{iT} \cdot \boldsymbol{k}_{jT}, \\
\hat{t}  =  (k_i-k_n)^2&=& -\frac{1}{z_n}(x_i\sqrt{s} p_T
e^{-y}+\frac{k_{iT}^2}{x_i\sqrt{s}} p_T e^{y}
-2\boldsymbol{k}_{iT}\cdot \boldsymbol{p}_{T}),\\
\hat{u} = (k_j-k_n)^2&=& -\frac{1}{z_n}(x_j\sqrt{s} p_T
e^{y}+\frac{k_{jT}^2}{x_j\sqrt{s}} p_T e^{-y}
-2\boldsymbol{k}_{jT} \cdot \boldsymbol{p}_{T}), \
\end{eqnarray}
and $\sqrt{s}$ is the center of mass energy of the incoming $p+p$ system,
$y$ is the rapidity of the produced hadron defined as $y=\frac{1}{2}
\textrm{ln}(E+p_L)/(E-p_L)$. To avoid the divergence of the partonic cross-sections when one
of the Mandelstam variables approaches to zero, we apply following
replacement
\begin{equation}
\hat{s} \to \hat{s}+ 2 \,\mu^2, ~~~~ \hat{t} \to \hat{t}- \mu^2,
~~~~~~ \hat{u} \to \hat{u}-\mu^2,
\end{equation}
with $\mu = 0.8$ GeV. This replacement will not change the partonic
cross-sections at large $\hat{s}$, $\hat{t}$ and $\hat{u}$. Another
advantage of this replacement is that the relation $\hat{s}+\hat{t}
+\hat{u} =0 $ is always satisfied.

The function $\mathcal{G}_p(k_{aT})$
describes the distribution of intrinsic transverse momenta carried
by partons. In phenomenological applications, a Gaussian form for
$\mathcal{G}_p(k_{T})$ is usually used:
\begin{equation}
\mathcal{G}_p(k_{T}) = \frac{\textrm{exp}(-k_{T}^2/\langle k^2_T
\rangle)}{\pi \langle k^2_T \rangle },\label{kt}
\end{equation}
where $\langle k^2_T \rangle $ is the square of the 2-dimensional RMS
width of the $k_T$ distribution for one parton, which is related to
the square of the 2-dimensional average of the absolute value of
$k_T$ of one parton through $\langle k^2_T
\rangle=4\langle k_T \rangle^2 /\pi$.  For simplicity, we
assume $\langle k^2_T \rangle $ to be independent of $Q^2$.
\begin{figure}
 \centerline{\includegraphics[height=.38\textheight]{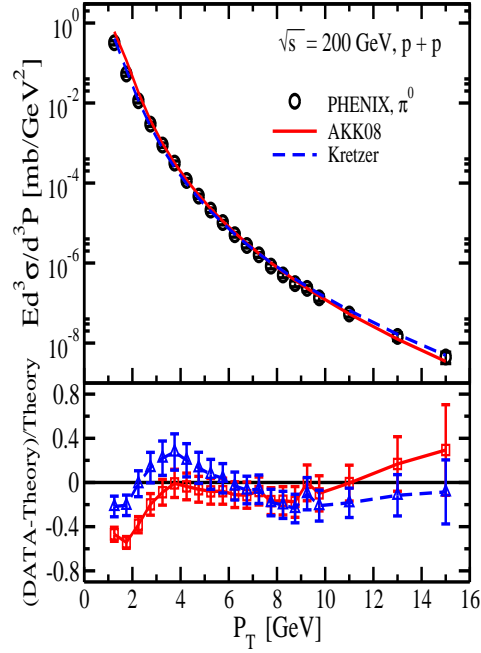}}
\caption{Cross
section of $p + p \to \pi^0 + X$ at RHIC energy $\sqrt{s}= 200$ GeV
for AKK08 and Kretzer fragmentation functions. The error bars are the
total uncertainties. Data are from the PHENIX collaboration
\cite{rhic2006}. \label{fig1}}
\end{figure}

\begin{figure}
\centerline{\includegraphics[height=.38\textheight]{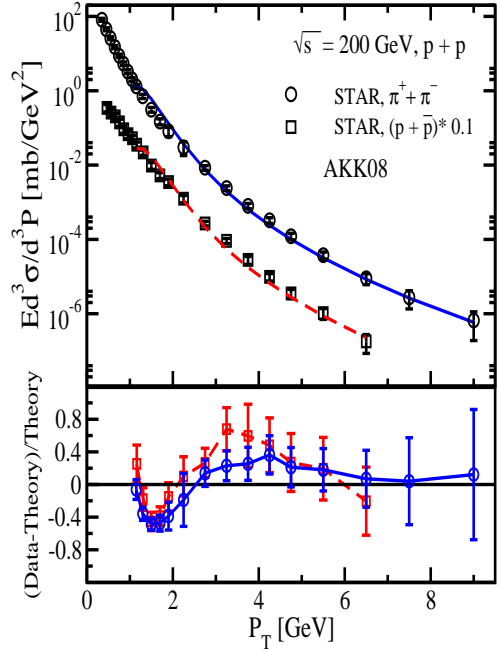}}
\caption{Cross
section of $p + p \to \pi^{+}(\pi^{-})+ X$ and $p + p \to
p(\overline{p})+ X$ at RHIC energy $\sqrt{s}= 200$ GeV. For theory
curves we use the parameter set with AKK08 fragmentation function, see
the text. The error bars are the total statistical and systematic
uncertainties.  Data are from the STAR collaboration
\cite{star-data}. \label{fig1-1}}
\end{figure}

\begin{figure}[!t]
 \centerline{\includegraphics[height=.27\textheight] {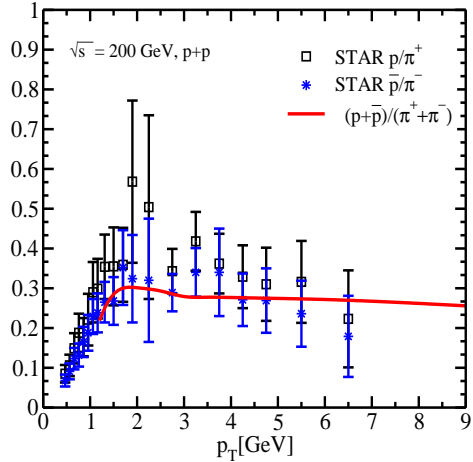}}
\caption{Ratio of $(p+\overline{p})/(\pi^{+}+\pi^{-})$ at RHIC energy $\sqrt{s}=200$ GeV for midrapidity with AKK08 FF. Data are for ratio of $\overline{p}/\pi^{-}$ and $p/\pi^{+}$
at midrapidity ($\mid y\mid<0.5$). The error bars are the total
statistical and systematic uncertainties. Data are taken from
Ref.~\cite{star-data}. \label{fig50}}
\end{figure}

With the factorization form shown in Eq.~(\ref{pptopi}), we calculate
the cross section of $p + p \to \pi^0 + X$ at RHIC for $\sqrt{s}= 200$
GeV. For the parton distribution $f_{i/p}(x,Q^2)$ used in our
calculations we adopt the newest MRST PDF ~\cite{mrst2006}. For the FF $D_{h/n}(z_n,Q^2)$, we first apply the
parametrization by Kretzer~\cite{kretzer2000}.  However,
Kretzer FFs do not give parton FFs to
protons. Therefore, we also consider AKK08 parametrization \cite{akk} of FFs
which give parton FFs to pions and $(p+\overline p)$. 
In both cases of calculations with Kretzer and AKK08 FFs, we will take $\langle k^2_T \rangle = 2 \,
\textrm{GeV}^2 $ and  $Q = p_T$ for the scale $Q$ of the hard process.
Notice also that different value of $\langle k^2_T \rangle$ has been
used in different approaches \cite{lhcf,ktc3}.  
We take the $K$-factor
$K=2.5, 1.5$ for the cases of Kretzer and AKK08 parametrization of FFs, respectively,
which gives a good approximation of the higher order contribution in
the $p_T$ region of interest.  With the settings mentioned above, we
are able to reproduce the pions ($\pi^0$, $\pi^{+}+\pi^{-}$) productions data from the STAR and
PHENIX collaborations \cite{rhic2006,star-data} for $p+p$
collisions. The discrepancy between theory and the data is within
40\%, see Figs.~(\ref{fig1},\ref{fig1-1}). The setting with AKK08 FF also
gives good predictions for protons productions in $p+p$ collisions in accordance
with data from the STAR collaboration \cite{star-data}, see
Fig.~(\ref{fig1-1}). 

In Fig.~(\ref{fig50}), we show experimental data from the STAR
collaboration for ratio of $\overline{p}/\pi^{-}$ and $p/\pi^{+}$ at
midrapidity. The ratio of
$(p+\overline{p})/(\pi^{+}+\pi^{-})$ at midrapidity obtained with AKK08 FF,
seems to be consistent with the experimental data for $\overline{p}/\pi^{-}$ and $p/\pi^{+}$.


\section{$p+A$ collisions and Cronin effect }
Multiple interactions of projectile partons in the target may proceed
coherently or incoherently. In the former case the multiple
interaction amplitude is a convolution of single scattering
amplitudes, and in the latter case one should convolute differential
cross sections, rather than amplitudes. The underlying mechanisms of
the multiple particle interactions and particle productions are
controlled by the coherence length \cite{abi1},
\beq
l_c\simeq \frac{\langle z \rangle  \sqrt{s}}{m_N p_T},\label{pt-coh}
\eeq
where $p_T$ is the transverse momentum of the fragmented hadron at midrapidity, and $m_N$ is the nucleon mass. 
For pion production, the average momentum fraction $\langle z \rangle$ in the FFs is about $0.4-0.6$ in the range of $2\leq p_T~(\text{GeV})\leq 8$.
For a coherence length which is shorter than the typical internucleon
separation $l_c \lesssim R_A$ (where $R_A$ denotes the nuclear radius), the projectile interacts
incoherently. At the RHIC energy, $\sqrt{s}=200\GeV$,
within intermediate $p_T$ we are almost in the transition regimes
between the short- and long-coherence limit for more central collisions, and at higher $p_T$ we are in the short-coherence
length (SCL) limit.

Here, we resort to a simple scheme assuming that the pQCD
factorization is valid at the RHIC energy at the midrapidity. 
We assume that the high-$p_T$ hadrons
are mainly originated from projectile's partons whose transverse
momentum is broadened by partons multi-scattering via gluons
exchange.  The broadening is computed in a parameter-free manner in
the color dipole approach. The single inclusive particle cross section
in minimum-biased $p(d)+A$ collisions can be written as
\begin{eqnarray}
\frac{d \sigma^{p(d)A\to h+X}}{dy d^2 \vec p_T d^2 \vec b} &=& 
\sum_{ijnl}\int dx_i dx_j d^2 k_{iT}d^2 k_{jT}
T_A(b) \,
f_{i/p(d)}(x_i,Q^2)\tilde{\mathcal{G}}_p(k_{iT},b,Q^2) \nonumber\\
&\times&\tilde f_{j/A}(x_j,b,Q^2) \mathcal{G}_p(k_{jT},Q^2)
K\, \frac{d \sigma}{d \hat{t}}(ij \to nl) \frac{D_{h/n}(z_n,Q^2)}{\pi z_n},
\label{fact1}\
\end{eqnarray}
where $ T_A(b)$ is the nuclear thickness function normalized to $\int
d^2b T_A(b)=A$. We will use the Woods-Saxon nuclear profile for $
T_A(b)$. $\tilde f_{j/A}(x_j,b,Q^2)$ denotes the parton
distribution function (Npdf) in the target nuclei (with atomic number $A$ and charge number $Z$) which can be parametrized in a factorizable form
\begin{eqnarray}
\tilde f_{j/A}(x_j,b,Q^2)&=&R_{j/A}(x, b, Q^2)\Big[\frac{Z}{A}f_{j/p}(x,Q^2)
+\left(1-\frac{Z}{A}\right)f_{j/n}(x,Q^2)\Big],\ \label{npdf}
\end{eqnarray}
in terms of parton distribution in a nucleon $f_{j/n}(x,Q^2)$ and the
nuclear modification function factor $R_{j/A}(x, b, Q^2)$.  We will
show the results of our calculation using three different Npdfs: EKS
\cite{eks} and HKN \cite{hkn} which are impact-parameter independent
but the $Q^2$-scale dependent via DGLAP evolution equation and
HIJING-new \cite{hij} which is impact-parameter dependent. 

Initial/final state broadening of the projectile/ejectile partons is
effectively taken into account via a modification of the primordial
transverse momentum distribution,
\beq
\tilde{\mathcal{G}}_p(k_{iT},b,Q^2)= \frac{d \sigma^{i=q}(qA\to qX) }{d^{2}\vec{k}_{iT}}(x,b), \label{broad} 
\eeq
where the transverse momentum distribution of partons after
propagation through nuclear matter of thickness $T_A(b)$ is subjected
to the broadening and computed in terms of the propagation of a
$q\bar{q}$ color dipole through the target nucleus \cite{boris0}
\begin{eqnarray}
\frac{d \sigma^{i=q}(qA\to qX) }{d^{2}\vec{p}_{T}}(x,b)&=&\frac{1}{(2\pi)^{2}}\int d^{2}\vec{r}_{1}d^{2}\vec{r}_{2}e^{i \vec{p}_{T}.(\vec{r}_{1}-\vec{r}_{2})}
\Omega^{q}_{in}(\vec{r}_{1},\vec{r}_{2})\nonumber\\
&\times& e^{-\frac{1}{2}\sigma_{q\bar{q}}(\vec{r}_1-\vec{r}_2,x) T_A(b)},\label{main}\nonumber\\
\end{eqnarray}
where $\Omega^{q}_{in}(\vec{r}_{1},\vec{r}_{2})$ is the density matrix which
describes the impact parameter distribution of the quark in the 
incident hadron, 
\begin{equation}
\Omega^{q}_{in}(\vec{r}_{1},\vec{r}_{2})=\frac{\langle k_{T}^{2}\rangle}{\pi} e^{-\frac{1}{2}(r_{1}^{2}+r_{2}^{2})\langle k_{T}^{2}\rangle},\label{pr}
\end{equation}
where $\langle k_{T}^{2}\rangle$ denotes the mean value of the parton
primordial transverse momentum squared. For a projectile gluon ($i=g$) we replace the $q\bar{q}$ dipole with a $gg$ one which can be written again in terms of $q\bar{q}$ dipole
via a Casimir factor $\sigma_{gg}=\frac{9}{4}\sigma_{q\bar{q}}$.  
The appearance of the dipole cross section in Eq.~(\ref{main}), is the
result of a product of the amplitude and the time-conjugated one,
which describe the quarks with different impact parameters. Clearly,
the object participating in the scattering is not a $q\bar{q}$ dipole
but rather a single colored parton.  To simplify the calculations we assumed that the
initial and final partons are the same, so the total nuclear
thickness $T_A(b)$ contributes to broadening.

We take for the dipole cross-section in Eq.~( \ref{main}) for the case of a projectile quark ($i=q$), the
popular saturation model of Golec-Biernat and W\"usthoff \cite{gbw}:
$\sigma_{q\bar{q}}(x,\vec{r})=\sigma_{0}\left(1-exp\left(-r^{2}/R_{0}^{2}\right)\right)$
where the parameters, fitted to DIS HERA data at small $x$, are given
by $\sigma_{0}=23.03$ mb, $R_{0}=0.4 \text{fm }\times
(x/x_{0})^{0.144}$, where $x_{0}=3.04\times 10^{-4}$. This
parametrization gives a good description of DIS data at
$x<0.01$ \cite{gbw}. 

The broadening in this scheme depends on the impact parameter
$b$, the transverse momentum of incident parton $k_{T}$, and also on
the target Bjorken $x$. Note that in Eq.~(\ref{fact1}), the summation
over index $i, j, n,l$ for quarks and gluons are different, not only
because the corresponding pdfs and FFs are different, but also because the
broadening of the projectile quarks and gluons are different.

The $K$-factor in
Eq.~(\ref{fact1}) and the parton primordial transverse momentum
squared $\langle k_{T}^{2}\rangle$ in
Eqs.~(\ref{fact1},\ref{pr}) are taken the same value fixed in
$p+p$ collisions. Therefore, all the phenomenological parameters in
the master equation (\ref{fact1}) are already fixed in reactions
different from $p(d)+A$ collisions. 

\begin{figure}[!t]
 \centerline{\includegraphics[height=.40\textheight] {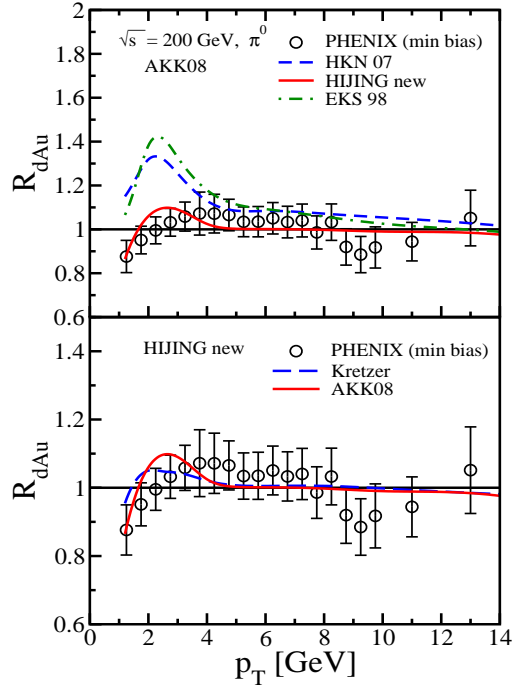}}
\caption{The mini-biased Cronin ratio $R$ for $\pi^0$ production at RHIC energy $\sqrt{s}=200$ GeV. All curves are at midrapidity. 
In the upper panel, we show $R_{dAu}$ with AKK08 FF for different nuclear
PDF. In the lower panel, we show $R_{dAu}$ with HIJING-new nuclear PDF
for different FFs. The normalization uncertainty on the $p+p$ reference of $9.7
\%$ is not included here. The error bars are the total statistical and
systematic uncertainties. Data are taken from
Ref.~\cite{rhic2006}. \label{fig2}}
\end{figure}

\begin{figure}[!t]
 \centerline{\includegraphics[height=.39\textheight] {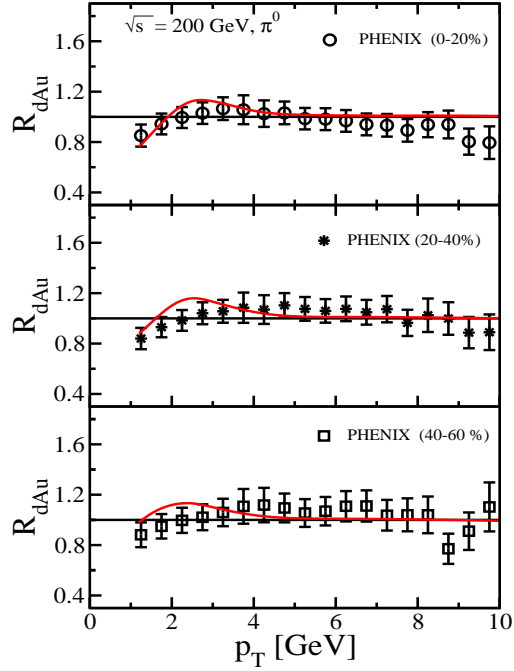}}
\caption{The Cronin ratio $R$ for $\pi^0$ production at RHIC energy $\sqrt{s}=200$ GeV at different centralities. All curves are at midrapidity 
with AKK08 FF and HIJING-new nuclear PDF.  Data
are taken from Ref.~\cite{rhic2006}. \label{fig3}}
\end{figure}

\begin{figure}[!t]
 \centerline{\includegraphics[height=.40\textheight] {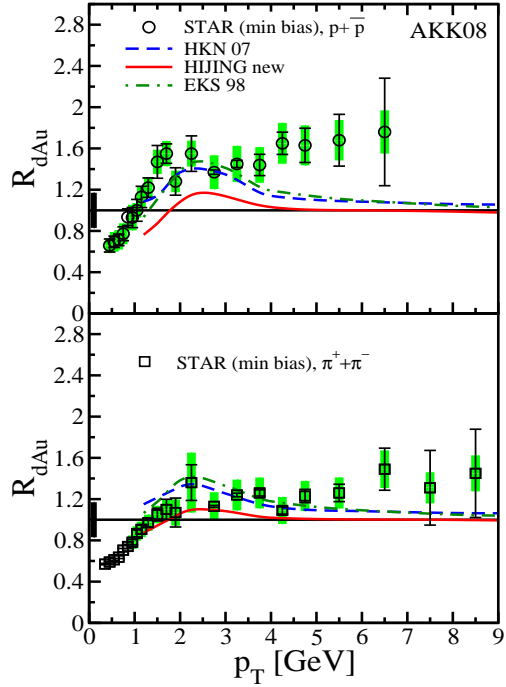}}
\caption{The mini-biased Cronin ratio $R$ for charged pions $\pi^{+}+\pi^{-}$ and baryons $p+\overline p$ productions at RHIC energy $\sqrt{s}=200$ GeV at midrapidity with AKK08 FF for various Npdfs. The experimental data are for $\mid y\mid<0.5$. The shaded bands around $1$ (of the order of $17\%$) corresponds to the error due
to uncertainties in estimating the number of binary collisions in minimum bias. 
Data are taken from
Ref.~\cite{star-data}. \label{fig4}}
\end{figure}


The nuclear modification factor $R_{pA}$ defined as ratio of $p+A$ to $p+p$ cross sections normalized to the average number of binary nucleon collisions,
\begin{equation}
R_{pA}=\frac{ \frac{d \sigma^{pA\to h+X}}{dy d^2 p_T}}{\langle N_{binary}\rangle\frac{d \sigma^{pp\to h+X}}{dy d^2 p_T}},
\end{equation}
where $\langle N_{binary}\rangle$ is the average number of geometrical
binary collisions which is calculated according to the Glauber model
\cite{gl} for different centrality.

In Fig.~(\ref{fig2}) (upper panel), we show mini-biased $R_{dAu}$ for
$\pi^0$ production at the RHIC energy for deuteron-Gold collisions with AKK08 FF for various
Npdfs. It is obvious that the uncertainty among various Npdfs at small
$p_T$ in the shadowing region leads to rather sizable different Cronin ratios. 
The position of the $R_{dAu}$ peak in $p_T$ does not seem to vary that much among various Npdfs. 

In Fig.~(\ref{fig2})
(lower panel), we show $R_{dAu}$ for $\pi^0$ production with
HIJING-new Npdf for different FFs. As one may expect both
considered fragmentation functions (Kretzer and AKK08 FFs) give very similar results for $R_{dAu}$, since we
assumed no medium modification for FFs.  Notice also that the
energy loss is less important at the RHIC energy for $d+Au$ collisions, although it might be important at lower energies at SPS
\cite{boris1,boris3}.

One should note that the nuclear interactions that lead to shadowing are also
the source of parton momentum broadening. We incorporate the shadowing
effect and the other nuclear medium modification of the partons through the Npdfs Eq.~(\ref{npdf}).  In principle, at
high-energy the $q\bar{q}$ dipole cross-section used for obtaining the
broadening is also subjected to the multi-Pomeron fusion effect in the
presence of nuclear medium. However, we have already incorporated the shadowing effect in the Npdf
and including the shadowing effect into the dipole cross-section may lead to double counting.
One should also note that the onset of gluon shadowing and its magnitude are still debatable, see Ref.~\cite{shd} and references therein.

\begin{figure}[!t]
 \centerline{\includegraphics[height=.27\textheight] {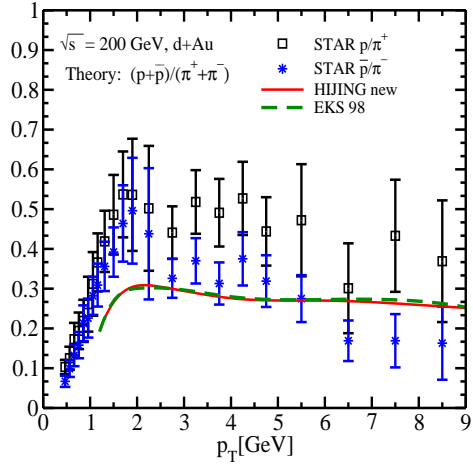}}
\caption{The mini-biased ratio of $(p+\overline{p})/(\pi^{+}+\pi^{-})$ for $d+Au$ collisions at RHIC energy $\sqrt{s}=200$ GeV for midrapidity with AKK08 FF and various Npdfs. 
Data are for ratio of $\overline{p}/\pi^{-}$ and $p/\pi^{+}$
at midrapidity ($\mid y\mid<0.5$) for minimum bias $d+Au$ collisions. The error bars are the total
statistical and systematic uncertainties. Data are taken from
Ref.~\cite{star-data}. \label{fig5}}
\end{figure}

In Fig.~(\ref{fig3}), we show the Cronin ratio at different
centralities at midrapidity. We used for all curves AKK08 FF and
HIJING-new Npdf. It is obvious that generally the centrality dependence
is rather weak in accordance with the PHENIX data \cite{rhic2006}. 

In Fig.~(\ref{fig4}), we show the mini-biased $R_{dAu}$ at midrapidity
for charged pions $\pi^{+}+\pi^{-}$ and baryons $p+\overline p$ at
midrapidity with AKK08 FF for various Npdfs. It is seen from
Figs.~(\ref{fig2},\ref{fig3},\ref{fig4}) that our results obtained
with HIJING-new Npdf agree with the Cronin data for both $\pi^0$ and
charged pions $\pi^{+}+\pi^{-}$. However, the same setting does not
seem to be in a good agreement with the Cronin data for baryons in
$p+A$ collisions although it gives a good description of baryons
cross-section in $p+p$ collisions. Taking the experimental data at face value,
the discrepancy between the theory and experimental data seems to persist up to $p_T=6.5$ GeV. 
A number of studies have recently found that recombination is more
important than fragmentation at small and moderate $p_T$ at midrapidity in heavy-ion collisions
at the RHIC energy $\sqrt{s}=200$ GeV \cite{fin1,finf2}. Therefore, a precise experimental measurement of the Cronin ratio for baryons at higher $p_T>7~\text{GeV}$ is essential  
in order to reveal the underlying mechanism of hadron production in the cold nuclear matter.  This deviation is also seen from
Fig.~(\ref{fig5}) where similar to Fig.~(\ref{fig50}), we plot
$(p+\overline{p})/(\pi^{+}+\pi^{-})$ ratio in $d+Au$ collisions at
midrapidity for various Npdfs. From Fig.~(\ref{fig5}), it is obvious
that baryons/pions ratio in $p+A$ collisions is not sensitive to the shadowing
effects in contrast to $R_{pA}$. To conclude, these discrepancies indicate that not entire Cronin effect in $p+A$ collisions comes from the
broadening due to initial partons multi-scatterings, and final-state effects which are not included here, are also important.

\begin{figure}[!t]
 \centerline{\includegraphics[height=.27\textheight] {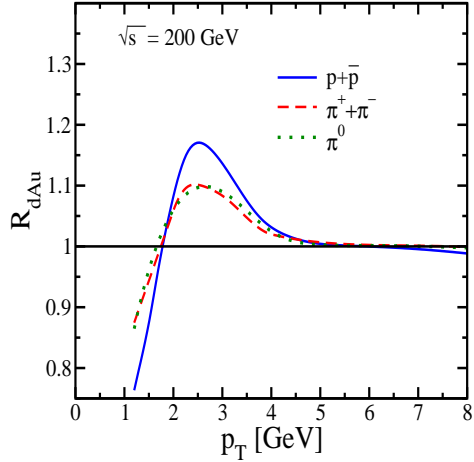}}
\caption{The mini-biased Cronin ratio for pions and protons at midrapidity for the RHIC energy with AKK08 FF and HIJING-new Npdf. \label{fig44}}
\end{figure}

In Fig.~(\ref{fig44}), we compare the mini-biased Cronin ratio for
pions and protons at midrapidity for the RHIC energy with AKK08 FF and
HIJING-new Npdf. It is obvious that the Cronin ratios for baryons and
mesons coming from initial-state effects are not significantly different.

\section{Summary and final remarks}
In this paper, we studied pions and baryons productions in $p+p$ and
$d+Au$ collisions at the RHIC energy. Having a successful description
of pions and protons productions in $p+p$ collisions, we showed that
the observed Cronin ratio for pions in $d+Au$ collisions can be fairly
described by transverse momentum broadening due to initial partons
multi-scatterings. But the same mechanism seems to underestimate the
observed Cronin ratio for protons in $d+Au$ collisions. We stress that
all phenomenological parameters in our model are fixed in reactions
different from $d+Au$ collisions.  This discrepancy might indicate
that initial state-effects (assuming that the fragmentation of hard
partons occurs outside of medium) might not be totally accountable for
the observed Cronin effect for baryons in $p+A$ collisions and the
separation of partons into two non-interacting components, soft and
hard, might be an oversimplification.  It is also possible that the
baryons productions mechanism in cold nuclear medium to be different from
pions productions.

A similar approach to this paper, was taken by Kopeliovich {\it et
al.} in Ref.~\cite{boris1} to calculate the mini-biased Cronin ratio
for $\pi^0$ . In this paper, we extended their study by calculating
the Cronin effect for the charged pions and protons at different
centralities.  One of the major difference between our study with
Ref.~\cite{boris1} is the way that the shadowing is included.  As we
already mentioned at moderate $p_T$ at the energy of RHIC, we are in
the transition regime between the short- and long-coherence limits.
Here, we assumed that the pQCD factorization is still valid in the
transition region, and we included the shadowing effect in the
conventional way via the nuclear parton distribution. We also studied
the effect of various available nuclear parton distributions. However,
in Ref.~\cite{boris1} no shadowing was included in Npdf. While the
Cronin ratio was obtained by a linear interpolation between two Cronin
ratios obtained in two different schemes of the short- and
long-coherence limits. We also tried their prescription in order to
investigate the pions and protons productions in $d+A$ collisions. We
found similar results to those obtained in our approach which leads us again to
the same conclusion that baryons and pions productions mechanism in
high-energy $p+A$ collisions may be different and final-state effects are important \cite{fin1,finf2}. Nevertheless, given rather
large experimental uncertainties further studies are needed in order to make a final verdict.


Notice that our prescription is not reliable at forward rapidities. The
reason is due to the importance of the valence quarks contribution which
are not incorporated in the color dipole picture. It might be tempting to assume that forward rapidity region
should be valid domain of our scheme since the Bjorken $x_2$ of target
is small, and the color-dipole approach which is based on
Pomeron-exchanges should be at work. However, one should note that at
this region $x_1\to 1$ and as a consequence the energy conservation
put a restricted constraint on the particle productions and therefore
valence quarks become very important \cite{break}.

\section*{Acknowledgments}
AHR would like to thank Boris Kopeliovich, Eugene Levin and Ivan Schmidt
for valuable discussions.
AHR wishes also to thank Hans-J\"urgen Pirner for asking a question
which led to this investigation. 
The authors are very grateful from Bedanga Mohanty and Simon Albino for useful communication. 
This work was supported in part by
Conicyt (Chile) Programa Bicentenario PSD-91-2006, Fondecyt (Chile)
grants 1070517 and Project of PBCT (Chile) No. ACT-028.


\end{document}